\documentclass{article}
\usepackage{spconf,amsmath,graphicx}
\usepackage{multirow,subcaption,graphics}
\usepackage{amssymb}
\usepackage{pifont}
\usepackage{diagbox}
\usepackage[table]{xcolor}
\newcommand{\xmark}{\ding{55}}%

\title{TOWARDS DOMAIN GENERALISATION IN ASR WITH ELITIST SAMPLING \\ AND ENSEMBLE KNOWLEDGE DISTILLATION}
\name{Rehan Ahmad, Md Asif Jalal, Muhammad Umar Farooq, Anna Ollerenshaw, Thomas Hain}
\address{Speech and Hearing Group, The University of Sheffield, UK}
\begin{document}
\ninept
\maketitle
\begin{abstract}
Knowledge distillation (KD) has widely been used for model compression and domain adaptation for speech applications. In the presence of multiple teachers, knowledge can easily be transferred to the student by averaging the models output. However, previous research shows that the student do not adapt well with such combination. This paper propose to use an elitist sampling strategy at the output of ensemble teacher models to select the best-decoded utterance generated by completely out-of-domain teacher models for generalizing unseen domain. The teacher models are trained on AMI, LibriSpeech and WSJ while the student is adapted for the Switchboard data. The results show that with the selection strategy based on the individual model’s posteriors the student model achieves a better WER compared to all the teachers and baselines with a minimum absolute improvement of about 8.4\%. Furthermore, an insights on the model adaptation with out-of-domain data has also been studied via correlation analysis.

\end{abstract}

\vspace{-4mm}
\section{Introduction}
\vspace{-2mm}
\label{sec:intro}

\let\thefootnote\relax\footnotetext{© 2023 IEEE.  Personal use of this material is permitted.  Permission from IEEE must be obtained for all other uses, in any current or future media, including reprinting/republishing this material for advertising or promotional purposes, creating new collective works, for resale or redistribution to servers or lists, or reuse of any copyrighted component of this work in other works.}

Knowledge distillation (KD) \cite{KD} is a popular technique to distill the knowledge from either single or multiple cumbersome teacher models to one student model. 
The knowledge from the teacher model is transferred at the output \cite{KD} or intermediate feature level \cite{romero2014fitnets}. A common technique is to transfer the knowledge using teachers' posterior distribution \cite{gou2021knowledge}. Usually, the student model is trained with the teacher's posteriors~(soft labels) along with the original training labels~(hard labels). Therefore, the total loss is the weighted sum of supervised and KD losses. KD is used for many tasks such as domain adaptation \cite{asami2017domain,meng2019domain,zhu2020domain}, domain generalisation \cite{wang2021embracing,kim2021domain, fang2021mosaicking}, and model compression \cite{chebotar2016distilling, kim2019knowledge, takashima2018investigation}. 


In automatic speech recognition, knowledge distillation is performed over either frame-level \cite{wong2016sequence} or sequence-level \cite{huang2018knowledge} span. Following the sequential nature of the ASR task, sequence-level transfer is shown to be a better approach in \cite{wong2016sequence}. Sequence level KD was first proposed in \cite{kim2016sequence}, where the teacher models provide sequence-level probability distribution over the whole sample space for better knowledge transfer. In the case of multiple teacher models, it is crucial to optimise the sampling strategy to obtain the best possible output. 
For example, \cite{gao2021distilling} proposed three types of selection strategies at the output of teachers, i.e. weighted, Top-1 and Top-K. The error metric is used as a selection criterion in all of these strategies. Therefore, the labelled data is assumed to be available. However, this approach is not helpful in those scenarios where the goal is to adapt unseen out-of-domain~(OOD) data. Hence, an unsupervised sampling strategy is necessary to reduce the uncertainty of multiple teacher outcomes and select the best output.


Furthermore, the underlying teacher model is important for the generalised representation of the acoustic samples. Various models such as BERT\cite{futami2020distilling}, BLSTMs \cite{8639629} with different context windows have been used in ASR knowledge distillation tasks. Self-supervised training paved the path to learn the general data representation through unsupervised pre-training. Such a model is trained with masked spans for generalised contextual latent representation of speech \cite{baevski2020wav2vec}. These self-supervised models have been observed to be quickly adaptable to new domains or cross-domain tasks \cite{hsu2021robust}.



The generalisation problem in KD is two-fold. The first problem is optimising the teacher model's learned representation, and the second is choosing the appropriate distribution for a student from the teacher model. These problems become challenging when dealing with OOD data between teacher and student. In this paper, the first problem is tackled with a pre-trained wav2vec teacher models where each model is fine-tuned with an in-domain corpora. The second problem is tackled with posterior based elitist sampling strategy which selects the best utterance decoded by the teachers. 
In summary, an ensemble of teacher models are trained on completely OOD data compared to the data which student needs to adapt. An inference is run on the unlabelled OOD data to generate the soft labels from the teachers. Finally, an elitist sampling strategy based on the output posteriors is used to select the best decoded utterance from the teachers to train a student model. More specifically, dataset from read speech: WSJ (LDC catalog LDC93S6A, LDC94S13A), LibriSpeech (LS) \cite{Panayotov2015} and meeting: AMI \cite{Carletta2006} are used to train three state-of-the-art teacher models. These models are used to decode the Switchboard (SB) \cite{Godfrey1992} corpus, and a student model is trained. 
The results show that with such selection on the ensemble of teacher outputs, the student model performs better compared to the baselines and all the individual teacher models.


\vspace{-3mm}
\section{distillation using ctc loss}
\vspace{-3mm}
\label{sec:kd}
In teacher-student training, knowledge is usually transferred from the teacher model in terms of the posterior distribution. The teacher model is already trained on given labelled dataset. While training the student, its output distribution is tried to become closer to the teacher's posterior distribution. This is usually achieved by using Kullback–Leibler (KL) divergence \cite{Joyce2011} loss as:

\vspace{-4mm}
\begin{equation}
\label{eq1}
    KL(\hat{y}_{tea} || \hat{y}_{stu}) = \sum_{i=1}^{N}\hat{y}_{tea}^i \log{\frac{\hat{y}_{tea}^i}{\hat{y}_{stu}^i}}
\end{equation}

where $\hat{y}_{tea}$ and $\hat{y}_{stu}$ represent the teacher and student output posteriors, and $N$ is the total number of examples. Frame-wise KL divergence is usually calculated to match the two distributions. The total loss for the student model is weighted sum of supervised and KL loss:

\vspace{-3mm}
\begin{equation}
\label{eq2}
    \mathcal{L}_{total} = \alpha \mathcal{L}_{sup} + (1-\alpha) \mathcal{L}_{KL}
\end{equation}
\vspace{-4mm}

In sequential tasks such as ASR, the studies show that knowledge can be transferred effectively via sequence-level information instead of frame-level. Therefore, CTC \cite{graves2012connectionist} loss has been used in this study. 
The teacher and student models for KD are based on wav2vec2 \cite{baevski2020wav2vec} pre-trained with libriVox (LV-60K). Dense neural network layers are connected at the output of wav2vec2 and CTC loss is applied. The raw input to the model is represented by sequence $ \mathbf{x} =[x_1, ... x_{T_n}] \in X $ with each sequence of length $T_n$ and $X$ represents all the training sequences. The wav2vec2 model outputs the context features represented by $ \mathbf{c} =[c_1, ... c_{T_n}]$.  Context vector $\mathbf{c}$ given to the fully connected dense layers produces the output represented by $ \mathbf{h} =[h_1, ... h_{T_y}]$, where $T_y$ is the output length with the vocabulary size of $G=[g_1,...g_z]$, where $z$ represents grapheme. Teacher models are trained with original labels using CTC loss as:

\vspace{-2mm}
\begin{equation}
\label{eq3}
    \mathcal{L}_{CTC} = -\sum_B \log p(\mathbf{y}|\mathbf{h})
\end{equation}
\vspace{-3mm}

where $\mathbf{y}$ is the output label sequence and $B=\{X,Y\}$ is the training dataset.


In a standard distillation setting, the student model is trained with the original (hard) labels and soft labels provided by the teacher model. The total loss for the student model is the weighted sum of hard and soft label losses as presented in equation \ref{eq2}. However, in our scenario, we assume that original labels of the target domain is not available.  
The OOD teacher models run the inference on the target data and provide the soft labels to train the student. Hence, the value of $\alpha$ in equation~\ref{eq2} becomes zero. 

For the distillation loss, we followed the work of \cite{takashima2018investigation,huang2018knowledge} which proposed to use sequence-level loss rather than frame-level for the ASR task. Therefore, $\mathcal{L}_{KL}$ in equation \ref{eq2} is replaced by soft CTC-KD loss. This is represented as follows:

\vspace{-2mm}
\begin{equation}
\label{eq4}
    \mathcal{L}_{CTC-KD} = - \sum_B p_{tea}(\mathbf{\hat{y}}|\mathbf{h_{tea}}) \log p_{stu}(\mathbf{\hat{y}}|\mathbf{h_{stu}})
\end{equation}
\vspace{-3mm}

where $p_{tea}(\mathbf{\hat{y}}|\mathbf{h_{tea}})$ and $p_{stu}(\mathbf{\hat{y}}|\mathbf{h_{stu}})$ are posteriors from teacher and student models. In this case, the teacher posteriors comes from the selection criteria from multiple teachers which is explained in Section \ref{sec: select}.

\vspace{-2mm}
\section{utterance selection strategies}
\vspace{-3mm}
\label{sec: select}


This work utilizes multiple teacher models to generalise the out-of-domain data for the student model. It is known that the ensemble models usually provide better predictions compared to the single model.  One of the most straightforward approaches to distill the knowledge from the teacher models is to average the frame-wise posterior of the teachers following the work of \cite{chebotar2016distilling,fukuda2017efficient} as: 

\vspace{-2mm}
\begin{equation}
\label{eq5}
    \mathbf{p_{tea}} = \frac{1}{K} \sum_k \mathbf{p_{k}}
\end{equation}
\vspace{-2mm}

were $K$ is the total number of teachers, and $\mathbf{p_k}$ is the posteriors of the whole utterance for teacher $k$. Resulting $\mathbf{p_{tea}}$ is the frame-wise average posteriors of each utterance. However, this strategy may not perform well because if some teachers perform poorly then the overall average become worse. Therefore, a sampling technique is devised which selects the best-decoded utterances from the teacher models.

In \cite{gao2021distilling}, authors proposed three distillation strategies, i.e. weighted, Top-1 and Top-K. The authors proposed to use error-rate as a metric of selection/weightage in each strategy. For example, in the weighted technique, the weights for each teacher are determined with the average error on the mini-batch, which can only be calculated if labelled data is present. Moreover, in their technique the student learns on the same data which teachers have already been trained on. However, this is not the case in our scenario, where the student learns on the data unseen by the teachers and without original labels. Eventually, this technique cannot be directly adapted. 

For the unlabelled data each teacher model may make different mistakes while transcribing the same utterance, and one teacher would perform better than the others. Therefore, we propose to select the utterance from that particular teacher model which best transcribes it. This could be achieved by ranking the decoded posteriors from the teachers. In our selection strategy, we propose to take the average of the posteriors of the whole utterance from each teacher as follows:

\vspace{-4mm}
\begin{equation}
\label{eq6}
    q_k = \frac{1}{T_y} \sum_{i=1}^{T_y} p_{ik}
\end{equation}
\vspace{-3mm}

where $T_y$ is the length of utterance and $k$ is the teacher. This average is calculated for each utterance which is decoded by the teachers. The best teacher is selected with maximum average posterior as:

\vspace{-4mm}
\begin{equation}
\label{eq7}
    b = \arg \max_k q_k
\end{equation}
\vspace{-3mm}

With such a selection strategy, each utterance is selected from the best performing teacher $b$, and its posteriors are used in equation (\ref{eq4}) to train the student model. The intuition behind the selection strategy is elitist, i.e., assuming if a group of graphemes projects the highest confidence as a group, they are also best fitted as individuals. The grapheme-phoneme correspondence will be optimal when we maximise the confidence score in graphemes as a group. If the posteriors are higher for each individual label, it refers to the model's confidence. Taking the average of all the frames would provide the model's confidence for the whole utterance. So the best utterance among the three models would have the highest average posterior.  

\vspace{-3mm}
\section{Internal Representation Correlation}
\vspace{-2mm}
\label{sec:svcca}
Understanding model learning from a core point of view is crucial to establishing the relationship between input acoustic data and the target labels. In ASR, the transcription often depends on the general spelling and linguistic perception of the transcribers' language dialect or country. Statistical correlation analysis is used to understand the relationships between representations in neural network layers \cite{ollerenshaw2021insights}. The activation outputs are extracted among layers or model outputs across model training/test, enabling observation of shared representations among representations.
SVCCA \cite{raghu2017svcca} uses singular value decomposition~(SVD) and canonical correlation~(CC) analysis for obtaining the shared relationships across neural representations. The algorithm attempts to find bases $v, \; s$ that maximise the correlation between two matrices (in this case, sets of neural layers) when the original matrices are projected onto the bases:

\vspace{-2mm}
\begin{equation}
\label{eq9}
    \frac{v^T\sum_{XY}s}{\sqrt{v^T\sum_{XX}v}\sqrt{s^T\sum_{YY}s}}
\end{equation}

Here $\sum_{XX}, \; \sum_{XY}, \; \sum_{YY}$ are the cross-covariance and covariance of the matrices respectively. The coefficiency of the correlations is assessed in a layer-wise approach for $\mathcal{N}$ data points, projected views of $l_1$ and $l_2$ are compiled, where $l_1=\{z_1^{l_1},...,z_{N_1}^{l_1}\}$ and $l_2=\{z_2^{l_2},...,z_{N_2}^{l_2}\}$. SVD is used to prune and preserve the top 99\% of representative dimensions, which forms subspaces $l_1' \subset l_1$ and $l_2' \subset l_2$. CC analysis is used to find vectors $v$ and $s$, which maximises the correlation $\rho$ between projections of $l_1'$ and $l_2'$. In this scenario, the value of $\rho$ correlates to the information encoded in the neural representations:

\vspace{-2mm}
\begin{equation}
\label{eq10}
    \rho = \frac{\langle v^T l_1',s^Tl_2'\rangle}{||v^T l_1'||\; ||s^T l_2'||}
\end{equation}

In this paper, two transformer (12 encoders, 12 decoder layers) models have been trained with the original Switchboard transcription and the teacher-ensemble generated pseudo transcription. The layer-wise embedding representations have been analysed with SVCCA and discussed in Section \ref{ref:sec:representation}.

\vspace{-3mm}
\section{experiments}
\vspace{-3mm}
\label{sec:exp}

\subsection{Data}
\label{data}
\vspace{-2mm}

Four different corpora have been used in the experiments comprising read and conversational speech. AMI, LS and WSJ are used to train the teacher models, while SwithBoard (SB) is for student. LS, WSJ and SB are recorded by US native English speakers, while AMI is recorded in three different institutes, including Edinburgh, Idiap and TNO, also covering non-native English speakers. Among these corpora, AMI and SB belong to the conversational speech and WSJ, and LS belongs to the read speech. Belonging to the same category does not reflect that they are similar. All of these corpora have different distributions in terms of recording conditions, speaking style etc.

The total duration of LS, WSJ, AMI and SB are 960h, 272h, 100h and 300h. As LS is the largest in duration, we only considered the 360h subset referred to as LS360. Moreover, all datasets have a sampling rate of 16Khz except SB. Hence, SB is up-sampled from 8KHz to 16KHz rate. To analyze the performance of the models, we considered the standard test sets from each corpus. LS-test-clean, LS-test-other are clean and noisy test sets of LS. One test from WSJ is considered, i.e. WSJ-Eval92-20K, which consists of 20K words vocabulary. From AMI, the full corpus ASR test set AMI-FCASC is taken. From SB, we considered two test sets; eval00-CallHome (LDC97S42) represented by CH and eval00-Switchboard (SB). The eval00 combines both of these CH and SB test sets. 

\vspace{-3mm}
\subsection{Experimental details}
\label{expdetail}
\vspace{-2mm}

The base model for teachers and student is wav2vec2-large pre-trained with LV-60k. The output of wav2vec2 is connected with two fully connected layers, and CTC loss is applied. 
During training, wav2vec2 is also fine-tuned for both teachers and the student. 
Moreover, wav2vec2 requires audio to be sampled at a 16KHz rate, so only SB audio is upsampled. 
The grapheme-based tokeniser is used for all the models.  


In the experiments, independent teacher models are first trained on AMI, WSJ and LS360 datasets. These models are evaluated on multiple test sets including in-domain and out-of-domain sets as listed in table \ref{tab:teacherwer}. 
This evaluation helps to understand the performance of the model on completely out-of-domain distribution to analyze cross-domain adaptation. The language model (LM) used in the decoding is a 3-gram model which is trained on the training transcripts of all the teacher models.

Teacher models are used to decode the SB data and based on the selection strategy discussed in section \ref{sec: select} the best decoded-utterance is selected to train the student model. While training the student model, it does not know the original training labels considering that they are not available. Hence, the student only generalises based on the posteriors from the teachers. Moreover, similar to the acoustic models the LM is also an out-of-domain model for the SB data. This is selected to maintain the consistency in this study to just evaluate the unseen domain adaptation. The results of the proposed technique are compared with individual teacher models' and two baseline. 
For the first baseline, frame-wise average of the posteriors of all the teachers are computed and resulting posteriors are used to train the student model. For the second baseline, we find the maximum frame-wise posteriors among the teachers and select it as resulting distribution. The resulting posteriors are then used to train the student. The frame-wise maximum value of posterior represents the confidence of the model for that particular label. Hence, frame-wise high confident soft-labels are selected. All of these results are discussed in Table \ref{tab:studentwer} and \ref{tab:studentwerLM}, where first baseline is represented by Tea\_avg (teachers averaging method) and second one as Fw\_max (frame-wise maximum method) .

\begin{table}[t]
    \centering
    \caption{WER(\%) evaluation of the teacher models on in-domain and out-of-domain test sets with and without LM. The three individual teacher models are trained on AMI, LS360 and WSJ datasets.}
    \label{tab:teacherwer}
    \begin{tabular}{p{2.5cm} |p{0.5cm} p{0.55cm} | p{0.5cm} p{0.55cm} | p{0.5cm} p{0.5cm}}
        \hline
        \hline
         
         & \multicolumn{2}{c|}{\bf AMI} & \multicolumn{2}{c|}{\bf LS360} & \multicolumn{2}{c}{\bf WSJ} \\
        \hline
        \diagbox{\bf Test sets}{\bf LM} & \xmark & \checkmark  & \xmark & \checkmark & \xmark & \checkmark \\
        \hline
        AMI-FCASC & \cellcolor{lightgray}15.85 & \cellcolor{lightgray}14.41 & 51.69 & 47.42 & 60.60 & 57.32 \\
        LS-test-clean & 14.61 & 12.41 & \cellcolor{lightgray}3.51 & \cellcolor{lightgray}3.04 & 10.14 & 8.39 \\
        LS-test-other & 29.69 & 26.32 & \cellcolor{lightgray}10.76 & \cellcolor{lightgray}9.51 & 27.19 & 24.43 \\
        WSJ-Eval92-20K & 15.87 & 13.11 & 9.60 & 7.14 & \cellcolor{lightgray}2.39 & \cellcolor{lightgray}1.47 \\

        eval00 & 52.04 & 49.21 & 45.86 & 42.59 & 66.43 & 64.20 \\
        \hline
        \hline
    \end{tabular}%
\vspace{-2mm}
\end{table}

\begin{table}[t]
    \centering
    \caption{Evaluation of WER(\%) of student and teacher models on swithboard test sets without language model.}
    \label{tab:studentwer}
    \begin{tabular}{p{0.7cm}|p{0.7cm}|p{0.7cm}|p{0.7cm}|p{0.9cm} p{0.9cm} p{0.9cm}}
        \hline
        \hline
         & \multicolumn{3}{c|}{\bf Teacher models} & \multicolumn{3}{c}{\bf Student models} \\
         & \multicolumn{3}{c|}{(w/o LM)} & \multicolumn{3}{c}{(w/o LM)} \\
         \hline
        & \multicolumn{3}{c|}{} & \bf Tea\_avg & \bf Fw\_max & \bf ours \\
        \hline
        \bf sets & \bf AMI & \bf LS360 & \bf WSJ & \multicolumn{3}{c}{\bf SWBD} \\
        \hline
        eval00 & 52.04 & 45.86 & 66.43 & 58.11 & 51.71 & \bf 37.38 \\
        CH & 56.74 & 50.55 & 73.48 & 62.68 & 54.63 & \bf 41.15 \\
        SB & 47.09 & 40.96 & 59.07 & 53.33 & 48.67 & \bf 33.45 \\
        \hline
        \hline
    \end{tabular}
\vspace{-5mm}
\end{table}

\begin{table}[t]
    \centering
    \caption{Evaluation of WER(\%) of student and teacher models on swithboard test sets with a 3-gram language model trained on AMI, LS360 and WSJ transcripts.}
    \label{tab:studentwerLM}
    \begin{tabular}{p{0.7cm}|p{0.7cm}|p{0.7cm}|p{0.7cm}|p{0.9cm} p{0.9cm} p{0.9cm}}
        \hline
        \hline
         & \multicolumn{3}{c|}{\bf Teacher models} & \multicolumn{3}{c}{\bf Student models} \\
         & \multicolumn{3}{c|}{(with OOD LM)} & \multicolumn{3}{c}{(with OOD LM)} \\
         \hline
        & \multicolumn{3}{c|}{} & \bf Tea\_avg & \bf Fw\_max & \bf ours \\
        \hline
        \bf sets & \bf AMI & \bf LS360 & \bf WSJ & \multicolumn{3}{c}{\bf SWBD} \\
        \hline
        eval00 & 49.21 & 42.59 & 64.20 & 46.53 & 38.09 & \bf 32.00 \\
        CH & 54.03 & 47.18 & 71.70 & 51.84 & 42.16 & \bf 35.72 \\
        SB & 44.17 & 37.80 & 56.37 & 40.99 & 33.84 & \bf 28.13 \\
        \hline
        \hline
    \end{tabular}
\vspace{-6mm}
\end{table}


\vspace{-4mm}
\section{Results \& Discussion}
\vspace{-3mm}
\label{sec:results}
The teacher models trained in individual corpora were initially evaluated on in-domain and out-of-fomain test sets. Table \ref{tab:teacherwer} shows the WER of the individual test set used for the evaluation. Each column (AMI, LS360 and WSJ) represents the model trained on that specific corpus and each test set is evaluated with and without LM.
It can be seen that each model produces state-of-the-art results on its own test sets represented by shaded cells. 
For both LS test sets the best cross-corpora WER was produced by WSJ and then AMI. This is due to the fact that WSJ is also read speech data and has native English speakers. However, both WSJ and LS have different recording and environmental conditions. Therefore, the WSJ model produced more than 2.5 times worse WER for LS-test-clean and LS-test-other with and without LM. On the other hand, AMI is a conversational speech and also has non-native English speakers. Additionally, the WSJ and LS are relatively more clean speech compared to AMI as it has many environmental and human-generated noises. Similar behaviour is being noticed for the WSJ-Eval92-20K test set. For the AMI test set, LS360 and WSJ produced worse results compared to the AMI model itself. Again the reason is the speaking styles and short spontaneous utterances present in the conversational speech. 
Finally, the eval00 test, which belongs to the SB data, is best transcribed by LS360. This is due to the large training data (360h) and more speakers compared to the AMI and WSJ. One can argue that the AMI should perform better due to the conversational speech. However, the AMI model is only trained on 86h and has both native and non-native English speakers. While SB only has native English speakers.

Table \ref{tab:studentwer} shows the WER of the individual teacher models and student models on SB test sets without LM. 
Among the student models the Tea\_avg do not seems be a good strategy for ensemble models as the WER is much worse compared two teachers (AMI and LS360) and also other student models. The main reason is that one of the teacher i.e. WSJ is performing worse, thus deteriorating the average of the teachers' posterior which eventually affected the student models training. Whereas, Fw\_max is better than majority of the teachers and Tea\_avg due to the reason that in this strategy maximum posterior is selected among the teachers which corresponds to the high confident prediction. The downside of this technique is that it affects the sequential nature of the output because it selects the posteriors from different teacher models.

Furthermore, other than the proposed the best model among the teachers and students which produced lowest WER for all the test sets is LS360. 
However, it cannot be concluded that the LS360 model decodes all the utterances of the SB better than AMI and WSJ to train the student. This is clearly depicted by evaluation of proposed student model which has lowest WER compared to all the teachers and baseline student models. 
For the CH test set, WER of proposed student model is 41.15\%, which is almost 9.4\% better than the LS360, 21.5\% better than Tea\_avg and 13.4\% better than Fw\_max in an absolute difference. 
Similarly, SB achieved minimum improvement of about 7.5\% compared to the LS360 model. 
Collectively, for eval00 the proposed student model is almost 8.4\% better than the best performing model LS360.

Moreover, Table \ref{tab:studentwerLM} show that better results are achieved when OOD LM is used for both teachers and students models. Compared to the results in Table \ref{tab:studentwer} the proposed student model is improved upto 5.3\% in absolute difference using the OOD LM making WER of eval00 to 32.00\%. It has also been observed that the results of Tea\_avg and Fw\_max models has improved significantly eventually making Fw\_max better than all the teachers and Tea\_avg. This shows that frame-wise posterior sampling seems to be a better candidate if decoding is applied with an LM.  
Overall, results show the significance of the ensemble of teachers and selection strategy to train the student. As the teacher models are trained on completely out-of-domain data, the knowledge can still be distilled effectively from ensemble models to adapt better student. Therefore, the student performs much better than the baselines and individual teachers.

\vspace{-3mm}
\subsection{Representation Analysis}
\label{ref:sec:representation}
\vspace{-2mm}


\vspace{-1mm}
\begin{figure}[h]
\vspace{-4mm}
    \centering
    \begin{subfigure}{0.49\linewidth}
        \includegraphics[height=0.8\linewidth,width=\linewidth]{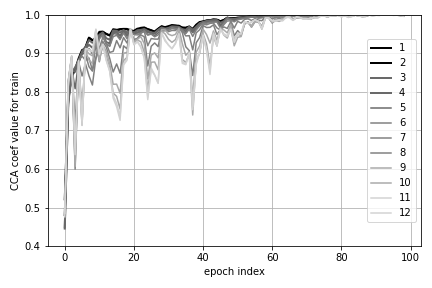}
    \end{subfigure}
    \begin{subfigure}{0.49\linewidth}
        \includegraphics[height=0.8\linewidth,width=\linewidth]{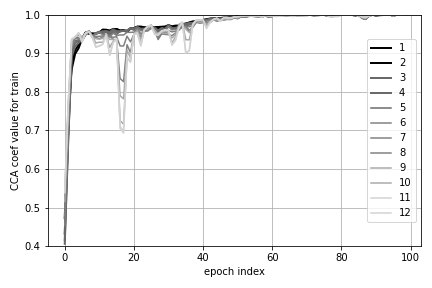}
    \end{subfigure}
    \caption{SVCCA correlation coefficients through time as performance converges within Transformer layers 1 to 12 model trained with original Switchboard transcripts (\textbf{left}) and Transformer layers 1 to 12 model trained with pseudo label transcripts (\textbf{right})}
    \label{fig:svcca_1}
\vspace{-4mm}
\end{figure}

\begin{figure}[h]
    \centering
    \begin{subfigure}{0.49\linewidth}
        \includegraphics[height=0.8\linewidth,width=\linewidth]{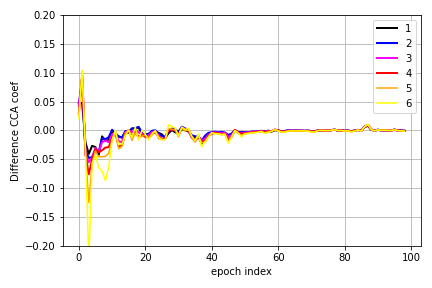}
    \end{subfigure}
    \begin{subfigure}{0.49\linewidth}
        \includegraphics[height=0.8\linewidth,width=\linewidth]{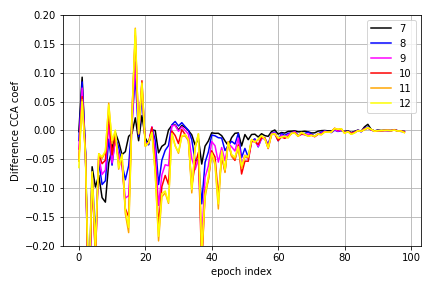}
    \end{subfigure}
    \caption{Difference in layer-wise SVCCA correlation between Switchboard model trained with original labels vs Switchboard model trained with pseudo labels through time within Transformer layers 1-6 (\textbf{left}) and layers 7-12( (\textbf{right}).}
    \label{fig:svcca_2}
\vspace{-6mm}
\end{figure}

\vspace{-1mm}
Figure \ref{fig:svcca_1} shows the correlation coefficiency through training of a model trained with Switchboard data and a model trained with Switchboard pseudo-labels. The convergence pattern shows during the training that the model layers first maximise the mutual information between the acoustic input and the layer latent representation and then minimise the mutual information between the layer representation and the target graphemes  \cite{shwartz2017opening, saphra2018understanding}. Figure \ref{fig:svcca_2} shows the difference in correlation coefficiency of the layers of a model trained with Switchboard original transcript and a model trained with Switchboard pseudo-label transcript. Figure \ref{fig:svcca_2} shows that the layers with closer proximity to acoustic input have very little to no difference among the Switchboard model trained with original transcripts vs the Switchboard model trained with pseudo label transcripts. However, the encoders' last three layers (10-12) show a considerable difference, which justifies the difference among the labels. It is safe to say that until the final few layers, both the models are learning similar representations. Thus the generated pseudo labels are not entirely different from the original label. 
Figures 1 \& 2 clearly show that, however, the middle layers learn similar representations, but the layers next to the target categorical distribution (graphemes) are sensitive and less stable than the original switchboard transcription model. This phenomenon is attributed to the confusion in pseudo label generation, mainly for acoustic boundary learning with hesitation before words and small interjections. Different pseudo labels were generated for identical utterances due to hesitation, laughing and other conversational nuances in those utterances, which made the model confused while training. The in-domain read-speech ensemble ASR models were not equipped with all these conversational speech nuances. These issues can be alleviated while pseudo label generation using language models similar to the unseen data domain.  

\vspace{-4mm}
\section{CONCLUSION}
\vspace{-3mm}
\label{conclu}
To generalise the unseen out-of-domain data, this paper proposed to use ensemble knowledge distillation with an elitist sampling technique. The sampling is performed based on output posteriors, and the best utterance is selected from the teachers to train a student model. With the proposed technique the results show that the student model generalises well on the out-of-domain data compared to the teacher models and baselines. Only out-of-domain language model is used to study the effectiveness of cross-domain acoustic model variability. Furthermore, intermediate neural representations are analysed across different models and layers to understand the relationship between acoustic data and transcription. In future work, one can improve the sampling strategy and pseudo label correction using iterative pseudo labeling and in-domain language models.

\vspace{-3mm}
\section{ACKNOWLEDGEMENTS}
\vspace{-3mm}
This work was conducted at the Voicebase/Liveperson Centre of Speech and Language Technology at the University off Sheffield which is supported by Liveperson, Inc. 

\bibliographystyle{IEEEbib}
\bibliography{refs}

\begin{thebibliography}{10}

\bibitem{KD}
G.~Hinton, O.~Vinyals, and J.~Dean,
\newblock ``Distilling the knowledge in a neural network,''
\newblock in {\em NIPS Deep Learning and Representation Learning Workshop},
  2015.

\bibitem{romero2014fitnets}
A.~Romero, N.~Ballas, S.~E. Kahou, A.~Chassang, C.~Gatta, and Y.~Bengio,
\newblock ``Fitnets: Hints for thin deep nets,''
\newblock {\em arXiv preprint arXiv:1412.6550}, 2014.

\bibitem{gou2021knowledge}
J.~Gou, B.~Yu, S.~J Maybank, and D.~Tao,
\newblock ``Knowledge distillation: A survey,''
\newblock {\em International Journal of Computer Vision}, vol. 129, no. 6, pp.
  1789--1819, 2021.

\bibitem{asami2017domain}
T.~Asami, R.~Masumura, Y.~Yamaguchi, H.~Masataki, and Y.~Aono,
\newblock ``Domain adaptation of dnn acoustic models using knowledge
  distillation,''
\newblock in {\em 2017 IEEE International Conference on Acoustics, Speech and
  Signal Processing (ICASSP)}. IEEE, 2017, pp. 5185--5189.

\bibitem{meng2019domain}
Z.~Meng, J.~Li, Y.~Gaur, and Y.~Gong,
\newblock ``Domain adaptation via teacher-student learning for end-to-end
  speech recognition,''
\newblock in {\em 2019 IEEE Automatic Speech Recognition and Understanding
  Workshop (ASRU)}. IEEE, 2019, pp. 268--275.

\bibitem{zhu2020domain}
H.~Zhu, J.~Zhao, Y.~Ren, L.~Wang, and P.~Zhang,
\newblock ``Domain adaptation using class similarity for robust speech
  recognition,''
\newblock {\em arXiv preprint arXiv:2011.02782}, 2020.

\bibitem{wang2021embracing}
Y.~Wang, H.~Li, L.-p. Chau, and A.~C Kot,
\newblock ``Embracing the dark knowledge: Domain generalization using
  regularized knowledge distillation,''
\newblock in {\em Proceedings of the 29th ACM International Conference on
  Multimedia}, 2021, pp. 2595--2604.

\bibitem{kim2021domain}
B.~Kim, S.~Yang, J.~Kim, and S.~Chang,
\newblock ``Domain generalization on efficient acoustic scene classification
  using residual normalization,''
\newblock {\em arXiv preprint arXiv:2111.06531}, 2021.

\bibitem{fang2021mosaicking}
G.~Fang, Y.~Bao, J.~Song, X.~Wang, D.~Xie, C.~Shen, and M.~Song,
\newblock ``Mosaicking to distill: Knowledge distillation from out-of-domain
  data,''
\newblock {\em Advances in Neural Information Processing Systems}, vol. 34, pp.
  11920--11932, 2021.

\bibitem{chebotar2016distilling}
Y.~Chebotar and A.~Waters,
\newblock ``Distilling knowledge from ensembles of neural networks for speech
  recognition.,''
\newblock in {\em Interspeech}, 2016, pp. 3439--3443.

\bibitem{kim2019knowledge}
H.-G. Kim, H.~Na, H.~Lee, J.~Lee, T.~G. Kang, M.-J. Lee, and Y.~S. Choi,
\newblock ``Knowledge distillation using output errors for self-attention
  end-to-end models,''
\newblock in {\em ICASSP 2019-2019 IEEE International Conference on Acoustics,
  Speech and Signal Processing (ICASSP)}. IEEE, 2019, pp. 6181--6185.

\bibitem{takashima2018investigation}
R.~Takashima, S.~Li, and H.~Kawai,
\newblock ``An investigation of a knowledge distillation method for ctc
  acoustic models,''
\newblock in {\em 2018 IEEE International Conference on Acoustics, Speech and
  Signal Processing (ICASSP)}. IEEE, 2018, pp. 5809--5813.

\bibitem{wong2016sequence}
J.~HM Wong and M.~Gales,
\newblock ``Sequence student-teacher training of deep neural networks,''
\newblock 2016.

\bibitem{huang2018knowledge}
M.~Huang, Y.~You, Z.~Chen, Y.~Qian, and K.~Yu,
\newblock ``Knowledge distillation for sequence model.,''
\newblock in {\em Interspeech}, 2018, pp. 3703--3707.

\bibitem{kim2016sequence}
Y.~Kim and A.~M Rush,
\newblock ``Sequence-level knowledge distillation,''
\newblock {\em arXiv preprint arXiv:1606.07947}, 2016.

\bibitem{gao2021distilling}
Y.~Gao, T.~Parcollet, and N.~D Lane,
\newblock ``Distilling knowledge from ensembles of acoustic models for joint
  ctc-attention end-to-end speech recognition,''
\newblock in {\em 2021 IEEE Automatic Speech Recognition and Understanding
  Workshop (ASRU)}. IEEE, 2021, pp. 138--145.

\bibitem{futami2020distilling}
H.~Futami, H.~Inaguma, S.~Ueno, M.~Mimura, S.~Sakai, and T.~Kawahara,
\newblock ``Distilling the knowledge of bert for sequence-to-sequence asr,''
\newblock {\em arXiv preprint arXiv:2008.03822}, 2020.

\bibitem{8639629}
G.~Kurata and K.~Audhkhasi,
\newblock ``Improved knowledge distillation from bi-directional to
  uni-directional lstm ctc for end-to-end speech recognition,''
\newblock in {\em 2018 IEEE Spoken Language Technology Workshop (SLT)}, 2018,
  pp. 411--417.

\bibitem{baevski2020wav2vec}
A.~Baevski, Y.~Zhou, A.~Mohamed, and M.~Auli,
\newblock ``wav2vec 2.0: A framework for self-supervised learning of speech
  representations,''
\newblock {\em Advances in Neural Information Processing Systems}, vol. 33, pp.
  12449--12460, 2020.

\bibitem{hsu2021robust}
W.-N. Hsu, A.~Sriram, A.~Baevski, T.~Likhomanenko, Q.~Xu, V.~Pratap, J.~Kahn,
  A.~Lee, R.~Collobert, G.~Synnaeve, and M.~Auli,
\newblock ``Robust wav2vec 2.0: Analyzing domain shift in self-supervised
  pre-training,''
\newblock {\em arXiv preprint arXiv:2104.01027}, 2021.

\bibitem{Panayotov2015}
V.~Panayotov, G.~Chen, D.~Povey, and S.~Khudanpur,
\newblock ``Librispeech: An asr corpus based on public domain audio books,''
\newblock in {\em IEEE ICASSP}, 2015, pp. 5206--5210.

\bibitem{Carletta2006}
J.~Carletta, S.~Ashby, S.~Bourban, M.~Flynn, M.~Guillemot, T.~Hain, J.~Kadlec,
  V.~Karaiskos, W.~Kraaij, M.~Kronenthal, G.~Lathoud, M.~Lincoln, A.~Lisowska,
  I.~McCowan, W.~Post, D.~Reidsma, and P.~Wellner,
\newblock ``The ami meeting corpus: A pre-announcement,''
\newblock in {\em Machine Learning for Multimodal Interaction}. 2006, pp.
  28--39, Springer Berlin Heidelberg.

\bibitem{Godfrey1992}
J.J. Godfrey, E.C. Holliman, and J.~McDaniel,
\newblock ``Switchboard: telephone speech corpus for research and
  development,''
\newblock in {\em IEEE ICASSP}, 1992, vol.~1, pp. 517--520 vol.1.

\bibitem{Joyce2011}
J.~M. Joyce,
\newblock {\em Kullback-Leibler Divergence}, pp. 720--722,
\newblock Springer Berlin Heidelberg, Berlin, Heidelberg, 2011.

\bibitem{graves2012connectionist}
A.~Graves,
\newblock ``Connectionist temporal classification,''
\newblock in {\em Supervised sequence labelling with recurrent neural
  networks}, pp. 61--93. Springer, 2012.

\bibitem{fukuda2017efficient}
T.~Fukuda, M.~Suzuki, G.~Kurata, S.~Thomas, J.~Cui, and B.~Ramabhadran,
\newblock ``Efficient knowledge distillation from an ensemble of teachers.,''
\newblock in {\em Interspeech}, 2017, pp. 3697--3701.

\bibitem{ollerenshaw2021insights}
A.~Ollerenshaw, M.~A. Jalal, and T.~Hain,
\newblock ``Insights on neural representations for end-to-end speech
  recognition,''
\newblock {\em Proc. Interspeech 2021}, pp. 4079--4083, 2021.

\bibitem{raghu2017svcca}
M.~Raghu, J.~Gilmer, J.~Yosinski, and J.~Sohl-Dickstein,
\newblock ``Svcca: Singular vector canonical correlation analysis for deep
  learning dynamics and interpretability,''
\newblock {\em Advances in neural information processing systems}, vol. 30,
  2017.

\bibitem{shwartz2017opening}
R.~Shwartz-Ziv and N.~Tishby,
\newblock ``Opening the black box of deep neural networks via information,''
\newblock {\em arXiv preprint arXiv:1703.00810}, 2017.

\bibitem{saphra2018understanding}
N.~Saphra and A.~Lopez,
\newblock ``Understanding learning dynamics of language models with svcca,''
\newblock {\em arXiv preprint arXiv:1811.00225}, 2018.

\end{thebibliography}

\end{document}